\documentclass[12pt,preprint]{aastex}
\def\sgra{\hbox{Sgr~A$^*$}}
 
\def\gsim{\;\lower4pt\hbox{${\buildrel\displaystyle >\over\sim}$}\;}
\def\lsim{\;\lower4pt\hbox{${\buildrel\displaystyle <\over\sim}$}\;}
\def\grls{\;\lower4pt\hbox{${\buildrel\displaystyle >\over <}$}\;}

\begin{document}
\title{An Imaging and Spectral Study of Ten X-Ray\\
Filaments around the Galactic Center}
\author{F. J. Lu\altaffilmark{1},
T. T. Yuan\altaffilmark{2},
Y.-Q. Lou\altaffilmark{3}}


\altaffiltext{1}{Key Laboratory of Particle Astrophysics,
Institute of High Energy Physics, Chinese Academy of Sciences
(CAS), Beijing 100049, P.R. China; lufj@ihep.ac.cn }
\altaffiltext{2}{Institute for Astronomy, University of Hawaii,
HI 96822, USA; yuantt@ifa.hawaii.edu}
\altaffiltext{3}{Physics Department and Tsinghua Center for
Astrophysics (THCA), Tsinghua University, Beijing 100084, P.R.
China; louyq@mail.tsinghua.edu.cn and lou@oddjob.uchicago.edu}

\begin{abstract}
We report the detection of 10 new X-ray filaments using the data
from the {\sl Chandra} X-ray satellite for the inner $6^{\prime}$
($\sim 15$ parsec) around the Galactic center (GC). All these X-ray
filaments are characterized by non-thermal energy spectra, and most
of them have point-like features at their heads that point inward. 
Fitted with the 
simple absorbed power-law model, the measured X-ray flux from an
individual filament in the $2-10$ keV band is $\sim 2.8\times10^{-14}$ to
$10^{-13}$ ergs cm$^{-2}$ s$^{-1}$ and the absorption-corrected
X-ray luminosity is $\sim 10^{32}-10^{33}$ ergs s$^{-1}$ at a
presumed distance of 8 kpc to the GC. We speculate the origin(s) of
these filaments by morphologies and by comparing their X-ray images
with the corresponding radio and infrared images. On the basis of
combined information available, we suspect that these X-ray
filaments might be pulsar wind nebulae (PWNe) associated with pulsars of
age $10^3 \sim 3\times 10^5$ yr. The fact that most of the filament 
tails point outward may further suggest a high velocity wind
blowing away form the GC.
 
\end{abstract}


\section{Introduction}

The Galactic center (GC) is the only place where we can observe
parsec details of various interaction in and around the Galactic
nucleus. Advances in this research frontier rely primarily on
observations at radio, infrared and X-ray wavelengths, because the
optical band suffers seriously from a considerable extinction with
an $Av\sim 30$ (e.g., Becklin, Matthews, Neugebauer \& Willner
1978). One of the most important discoveries in the GC region is
perhaps the presence of many structured non-thermal radio
filaments (NTFs) (e.g., Yusef-Zadeh, Morris, \& Chance 1984;
Morris \& Serabyn 1996; LaRosa, Kassim, Lazio, \& Hyman 2000).
While these non-thermal radio filaments have been intensively
studied, their origins and implications on the underlying physical
processes around the GC region remain largely unclear. In the
current analysis of X-ray filaments around the GC region, these
X-ray emitting particles are usually expected to be fairly close
to their acceleration zone and evolve very rapidly in time. Thus,
the X-ray study of the same region would be essential to probe the
origin of these energetic particles. The {\sl Chandra} Galactic
center Survey (CGS), with its unprecedented high spatial
resolution of $\sim 0.5''$ and moderate spectroscopy capability,
has already revealed remarkable X-ray structures (including
thousands of X-ray bright point sources and some filaments, as
well as clumps of diffuse emission) within the central $200$ pc
of our Galaxy (e.g., Wang, Lu, \& Lang 2002a). In this paper,
we mainly concentrate on the nature of those thread-like linear
structures or filaments as observed in X-ray bands.

Up to this point within $15^\prime$ ($\sim 37$ pc at 8 kpc) from
\hbox{Sgr~A$^*$} where a $\sim 4\times 10^6M_{\odot}$ black hole
resides inside a compact region of less than $\sim 1$AU (e.g.,
Shen et al. 2005), 5 X-ray filaments have been studied in details
(see also Table 1). For example, G359.95-0.04, a comet-like
filament at $\sim 0.32$ pc north of \hbox{Sgr~A$^*$}, is thought
to be a ram-pressure confined pulsar wind nebula (PWN) (Wang et
al. 2006).
Another prominent filament, G359.89-0.08 (SgrA-E), at $\sim 7$ pc
southeast of \hbox{Sgr~A$^*$}, was first noticed by Sakano et al.
(2003) and an interpretation of a possible PWN origin was discussed
in details by Lu, Wang \& Lang (2003). An alternative picture of
G359.89-0.08 as a source of synchrotron emission from relativistic
particles accelerated by a shock wave of W28 SNR was suggested
recently by Yusef-Zadeh et al. (2005); in that same paper, they also
detected a new X-ray filament G359.90-0.06, which coincides
spatially with a radio filament at $\sim 5.8$ pc southwest of
\hbox{Sgr~A$^*$}. They explored the mechanism of an inverse Compton
scattering (ICS) for the X-ray emission of G359.90-0.06. Another 2
filaments were found in more extensive regions. G0.13-0.11 in the
Radio Arc region was first reported by Wang, Lu \& Lang (2003) and
was suspected to be also a PWN. Of particular interest is the X-ray
filament G359.54+0.18; it associates with only one strand of the
obviously bifurcated radio threads (e.g., Yusef-Zadeh et al. 2005;
Lu et al. 2003).
A common feature shared in the X-ray energy spectra of these filaments
is that they all appear to be non-thermal. It should be noted however
that any thermal component to these sources would likely be completely
absorbed and unobservable, given the high foreground column density.

In these earlier investigations, pulsar wind nebulae (PWNs) and
supernova remnants (SNRs) seem to offer natural explanations for
the appearance of such X-ray filaments. Indeed, it is believed
that a considerable number of supernovae should have happened in
the GC region
(e.g., Figer et al. 1999, 2004; Wang et al. 2006).
One would naturally expect to find some of their end-products such
as pulsars and SNRs in the GC region. However, no radio pulsars
have yet been found within $\sim 1^\circ$ of the GC (Wang et al.
2006).
This might be caused by difficulties in radio observations
(Cordes \& Lazio 1997; Johnston et al. 2006).
Seeking observational clues in X-ray bands might shed new light to
the search for radio pulsars embedded in the GC region.

Another tempting idea is to use these X-ray filaments as potential
tracers for the magnetic field and gas dynamics around the GC region,
since the magnetic fields should have played a significant role in 
producing such
non-thermal spectra and the thread-like shapes of these filaments and
the gas motion is usually coupled with the magnetic fields. (e.g., Chevalier 1992;
Boldyrev \& Yusef-Zadeh 2006).
Magnetic fields exist on all scales of the Galaxy as well as in other
spiral galaxies and generally trace out spiral patterns on large scales
(e.g., Beck \& Hoernes 1996; Fan \& Lou 1996; Zweibel \& Heiles 1997;
Wielebinski 2005; Ferriere 2001, 2007),
and great progress has been made in measuring them and inferring
their influence by various means. For example, the observed
high-energy cosmic ray anisotropy at a few times $10^{-3}$ (e.g.,
Amenomori et al. 2006) might be physically related to large-scale
structures of Galactic magnetic fields and inhomogeneous cosmic ray
source distribution. Using diffuse synchrotron radio emissions at 74
and 330 MHz frequencies produced by relativistic cosmic-ray
electrons and the magnetic field around the Galactic center, LaRosa
et al. (2005) inferred a weak magnetic field of order of $\sim 10
\mu$G on size scales $\gsim 125''$ based on a minimum-energy
analysis. This is about 2 orders of magnitude lower than $\sim 1$ mG
usually estimated for the GC region. Very recently, Cowin \& Morris
(2007) argued that the assumption of LaRosa et al. (2005) that the
magnetic field and cosmic rays are in a minimum-energy state across
this region is unlikely to be valid. According to their model
estimates, the mean magnetic field is at least 100 microgauss on a
scale of several hundred parsecs and peaks at approximately 500
microgauss at the center of the diffuse nonthermal source (DNS).
This is an important issue to be settled for the GC magnetic
environment. Most of the GC radio NTFs are found to be perpendicular
to the Galactic plane, implying a local poloidal magnetic fields of
about milli-gauss strengths (e.g., Yusef-Zadeh \& Morris 1987). However, recent
observations revealed that the GC magnetic field may be more
complex than a simple globally ordered dipolar field (e.g.,
LaRosa et al. 2004; Nord et al. 2004). It might be possible for
X-ray NTFs to also provide clues of the configuration of the
local magnetic field as well as the interaction between it and 
the ambient gas flow. 

We report morphological and spectral properties of another 10 newly
discovered X-ray filaments within a region of $6^\prime .5 \times
6^\prime .5$ surrounding the \hbox{Sgr~A$^*$} (roughly corresponding
to a projected sky area of $\sim 15$ pc by $\sim 15$ pc at a
presumed distance of 8 kpc to the GC). Their plausible physical
origins are discussed in section 4. All error bars in the X-ray
spectrum parameter measurements are at the 90\% confidence level,
and we express the fitted parameters in the format of $y\ (y_l,\
y_u)$, where $y$ is the best fit value while $y_l$ and $y_u$ are the
lower and upper limits of the 90\% confidence interval,
respectively. For all the images in the paper, north is up and east
is to the left. We shall adopt a distance of 8 kpc from the solar
system to the GC throughout this paper. We note that during the
review process of this manuscript, Muno et al. (2007) submitted a
paper also on X-ray filaments around the Galactic center.

\section{Satellite Observations and Data Reduction}

This work takes advantage of a large number of observations by the
Advanced CCD Imaging Spectrometer (ACIS-I) onboard of {\sl
Chandra} X-ray satellite aimed at the \hbox{Sgr~A$^*$} (e.g.,
Baganoff et al. 2003).
We utilize the archival data available by September 2005,
including the eleven observations taken before May 3, 2002, as
listed by Park et al. (2005) (1561b excluded), and the three more
recent observations (Obs. ID 3549, 4683, and 4684) taken on June
19, 2003 and July 5 and 6, 2004. The total resultant exposure time
of the included 14 observations is $\sim$ 760 ks. Following the
standard event reprocessing procedures from the package of Chandra
Interactive Analysis of Observations (CIAO) (version 3.2.2), we
first processed individual observations, including the correction
for charge transfer inefficiency, bad-pixel-removing, and
light-curve cleaning etc. We then co-add the data from the 14
observations by the sky coordinates to generate the composite data
set. The detailed data preparation procedures are described in
Wang et al. (2006).

When placed on axis at the focal plane of the grazing-incidence X-ray
mirrors, the imaging resolution (FWHM) is determined primarily by the
pixel size of the CCDs of 0$\farcs$492. The CCDs also measure the
energies of incident photons with a resolution of $50-300$ eV
(depending on photon energy and distance from the read-out node)
(e.g., Baganoff et al. 2003).
We used the X-ray data in $0.5-8$ keV energy band for the spectrum
study. The CCD frames are read out every 3.2 s, which provides the
nominal time resolution of the X-ray data (e.g., Muno et al. 2006).
Here, we focus on the inner $\sim$ $6\farcm5$ region centered
around the \hbox{Sgr~A$^*$} where the point spread function (PSF)
broadening is insignificant.

\section{Data Analysis and Results on X-Ray Filaments}

The source region framed for extracting the spectrum of each source
is chosen in such a way that it is small enough to minimize
contaminations from the surroundings and yet large enough to contain
the bulk of filament emission. For those X-ray filaments with a
sufficient number of photons for statistics, we divide them into
sub-regions to look for the spectral evolution along the filamentary
structure. The background spectrum is extracted from one or several
regions in the environs of the relevant filament under
consideration. For each filament, we have used different background
regions to test their effects on the final spectra. Although the
best fit parameters do vary slightly with different backgrounds,
this variation is much smaller than the uncertainties of the current
data statistics. Most of the X-ray spectra are modelled by the
absorbed power-law continuum ($phabs(po)$ for short hereafter) in
{\sl XSPEC}. The Galactic coordinate of a filament is given by the
position of its brightest part. Incidentally, the size of the
filament given in this paper should only be regarded as a rough
estimate.

Fig. 1 gives a panoramic view of the X-ray filaments
in the GC region studied here. \sgra\ is buried deep in the lower
right region where the emission is saturated in Fig. 1. We studied
the images of each X-ray filament in details and the results are
shown in the left columns of Fig. 2 and
Fig. 3. From the X-ray contours, we can see that
there are signs for the existence of point-like sources in some
filaments. The right columns of Fig. 2 and
Fig. 3 show the $0.5-8$ keV counts image created from
the composite event file of the 14 {\sl Chandra} observations as
mentioned in the last section. The rectangular boxes drawn on the
images represent the source regions (including the sub-regions used
for studying spectral evolution along a filament).
Fig. 4 shows the final spectra of the ten filaments
extracted from the ``entire'' source regions. All of them can be
well fitted with the $phabs(po)$ model, implying that the X-ray
emissions we detected are dominated by the nonthermal component. It
should be noted that if these sources emit any thermal components,
they would likely be completely absorbed and unobservable. In
Fig. 5, we give the joint-fit spectra extracted from
the sub-regions of filaments F1, F3, F6 and F10, whose photon counts
are large enough to allow for a comparison of the sub-region
spectra. The spectrum fitting results are summarized in
Table 1.

We have also sought for the counterparts of X-ray filaments in H, K
infrared bands and in 20 cm radio band. The infrared images were
produced by the 2 Micro All Sky Survey project and were downloaded
from the {\it skyview} website.\footnote{The website address is
http://skyview.gsfc.nasa.gov/cgi-bin/query.pl\ .} The radio data
were acquired by the Very Large Array (VLA) on July 23, 2001, with a
beam size of 15$\arcsec\times15\arcsec$ (Lang et al. 2007). No
obvious filamentary structures were found (with an
upper limit of 0.12 Jansky/beam), except for filament F10,
which coincides with a statistically significant radio filament. In
addition to the fitting parameters for filaments summarized in
Table 1, we now describe properties of each X-ray
filament below.

\noindent {\bf Filament 1 (F1): G359.936-0.038}

Filament F1 lies $\sim 41\arcsec$ ($\sim 1.6$ pc) southwest of
\hbox{Sgr~A$^*$}. The brightest part of F1 is near the Celestial
coordinates {\it (l, b)}=($359^{\circ}.936$, $-0^{\circ}.038$). The
elongation occurs almost in the exact north-south direction with a
point-like source residing at the head in the northeast part. It has
a rough length of $\sim 17\arcsec$ and a width of $\sim 5\arcsec$.
Its spectra can be well fitted by $phabs(po)$. We divided the
filament into the head and tail regions, and fitted their respective
spectra. Although the  the best-fit
$\Gamma_{tail}$ appears twice bigger than $\Gamma_{head}$, the error
bars are larger than the difference. Therefore, the current data are
not good enough to constrain the spectral evolution of F1.

\noindent {\bf Filament 2 (F2): G359.934-0.037}

Filament F2 has a size of $3\arcsec\times 9\arcsec$. The most
concentrated part of F2 is near the Celestial coordinates {\it (l,
b)} = ($359\fdg 934$, $-0\fdg 037$), which is $\sim 48\arcsec$
($\sim 2~{\rm pc}$) southwest of \sgra. The elongation is along the
northeast-southwest direction with the head in the northeast end.
Similar to F1, there is no obvious emission line and a simple
power-law gives a fairly satisfactory fit.

\noindent {\bf Filament 3 (F3): G359.965-0.053}

At $1\farcm3$ ($\sim$3 pc) northeast of \sgra\ lies filament F3
with a size of $4\arcsec\times 12\arcsec$; the elongation is
from northeast (NE) to southwest (SW). This filament was first
mentioned by Baganoff et al. (2003)
as a ``curious linear feature". They reported its existence in the
$1.5-3$ keV and $3-6$ keV bands, but not in the $6-7$ keV band.
Our analysis also shows that the bulk of F3 emission comes from
photons of energies below 6 keV. The X-ray morphology clearly
shows that F3 contains two concentrations. In order to study their
spectral evolution, we divide it into the ``head'' (the NE
concentration with stronger emission), ``middle'' (between NE and
SW), and ``tail'' (the SW concentration with weaker emission)
regions. There appears a spectral steepening from the ``tail'' to
``head''. However, since no obvious point-like source is detected
in the ``head'' region, the ``head'' to the ``tail'' direction
does not necessariely mean the electron flow direction, as will
be further discussed in section 4.3.

The two-blob morphology is somewhat unexpected and one might wonder
whether the three regions we choose are in fact coherent parts of
the same source. Actually, the spectral results may cast doubt on
the coherence of the ``head'' and ``tail''. First, the $N_H$ value
in Table 1 shows that the ``head'' region has a
hydrogen column density of $1.9(1.3, 2.6)\times 10^{23}$ cm$^{-2}$,
while the ``tail'' region has a $N_H$ value of $0.7(0.3, 1.1)\times
10^{23}$ cm$^{-2}$, which might suggests that the ``tail'' could be
nearer to us than the ``head" is. Second, the spectrum of the
``tail" shows slight signs of 6.7 keV emission line, which is not
found in the ``head" and ``middle" regions.

However, we find that these spectral ``differences'' are most likely
not caused by the emitter proper. The higher $N_H$ to the ``head"
region is quite probably due to a small molecular cloud in this
direction (e.g., Coil \& Ho 1999, 2000). Fig. 2 of Coil \& Ho
(2000) shows that the column density of the NH3 cloud in the
``head'' region is about 1/10 of that of the streamer at
R.A.$_{1950}=19 42 28$, Dec$_{1950}=-$29 02 30, while the equivalent
$N_H$ of this streamer is $\sim 1.4\times 10^{24}$ cm$^{-2}$ (Coil
\& Ho 1999). So $N_H$ of the small NH3 cloud is $\sim 1.4\times
10^{23}$ cm$^{-2}$, comparable to the $N_H$ difference between the
``head'' and ``tail'' spectra. The slight 6.7 keV emission line
feature in the spectrum of the ``tail'' is likely due to an
insufficient subtraction of the background emission. For comparison,
we extract the spectrum of the dense diffuse emission near the
``tail" region and do find that it has a fairly strong 6.7 keV iron
line (see the upper panel of Fig. 6). Therefore, F3
is most likely one source object in space.

\noindent {\bf Filament 4 (F4): G359.964-0.056}

Filament F4 is close to filament F3 in sky projection. The
elongation occurs in the northeast-southwest direction. It has a
size of $4\arcsec\times 12\arcsec$. The spectrum also appears to be
non-thermal.

\noindent {\bf Filament 5 (F5): G359.959-0.028}

This filament lies $1.4^{\prime}$ ($\sim 3.3~{\rm pc}$) northwest of
\sgra, with a length of $\sim 10\arcsec$ and a width of $\sim
3\arcsec$. The elongation is oriented along the northwest-southeast
direction for fitting parameters).
There exists a point source near the southeast end of F5 (separation
$\sim 2\arcsec$); however, this is very likely a chance spatial
coincidence and thus we do not include it in the source region.

\noindent {\bf Filament 6 (F6): G359.969-0.038}

Filament F6 lies $\sim 1.6^{\prime}$ ($\sim 4~{\rm pc}$) northeast
of \sgra. It has a width of $\sim 5\arcsec$ and a length of $\sim 16
\arcsec$. The elongation occurs in the northeast-southwest
direction, and a point-like source resides at the southwest end.
Along the elongation, we divide F6 into the head (SW) and  tail (NE)
parts. The spectra of these two parts are the same within error bars.

\noindent {\bf Filament 7 (F7): G359.920-0.029}

Filament F7 also lies $\sim 1.3^{\prime}$ ($\sim 3~{\rm pc}$)
southwest of \sgra, with a size of $\sim 4\arcsec\times 15\arcsec$.
The elongation is along the northwest-southeast direction fairly
close to the east-west direction. Since there are no X-ray photons
below 4 keV, the statistics are extremely poor for this filament and
therefore, the energy spectrum cannot be fitted to distinguish any
models. Nevertheless, we still fit it with an absorbed power-law
model.

\noindent {\bf Filament 8 (F8): G359.970-0.009}

Filament F8 is located $\sim 2.7^{\prime}$ ($\sim 6~{\rm pc}$)
northwest of \sgra. The elongation is along the northwest-southeast
direction. It has a width of $\sim 4\arcsec$ and a length of $\sim
9\arcsec$. The energy spectrum of filament F8 also appears to be
non-thermal.

\noindent {\bf Filament 9 (F9): G359.974-0.00}

Filament F9 is $\sim 3\farcm3$ ($\sim 8~{\rm pc}$) northwest of
\sgra.  Its elongation is almost parallel to those of filaments F5
and F8. It has a width of $\sim 4\arcsec$ and a length of $\sim
7\arcsec$. The energy spectrum of filament F9 also appears to be
non-thermal.

\noindent {\bf Filament 10 (F10): G0.029-0.06}

This filament lies $\sim 5.1^{\prime}$ ($\sim 12~{\rm pc}$)
northeast of \sgra. With a width of $\sim 6\arcsec$ and a length of
$\sim 46\arcsec$, it is the longest X-ray filament so far identified
within the central 15 pc of the Galaxy center. The energy spectrum
of filament F10 is rather flat and non-thermal,
with the column density
of a typical value for the absorption around the GC.
To see the photon index evolution along the filament, we divide the
filament into three parts: ``head'' (i.e. where the assumed
point-source resides), ``middle'', and ``tail'' parts. We then
conducted a joint fit for the energy spectra of these three regions.
As listed in Table 1, there appears a spectral
steepening from the ``head'' to ``tail''.  No IR counterparts are
found for filament F10. But quite interestingly, there is a 20cm
radio filament more or less coincident with F10 spatially;
implications of this spatial coincidence will be discussed in $\S
4.2$.

\section{Discussion}

\subsection{\label{subsec:origins}Properties
and possible origins of GC X-ray filaments }

While X-ray photon numbers may not be high enough to constrain the
exact shapes of the energy spectra, all spectra appear featureless
except for filament F3 showing weak iron line features (see Section
3.3) and can be well fitted with power-law models. Fitting of some
of these energy spectra with thermal emission models is acceptable
statistically; nevertheless, this always gives quite high
temperatures, i.e., $>10$ keV. We therefore incline to the view that
X-ray emissions from these filaments are non-thermal in nature.

Table 2 sums up the inferred parameters for
these 15 non-thermal X-ray filaments in the inner $15^{\prime}$
around the GC. In addition to the 10 filaments studied in this
paper, we also include the other 5 filaments, namely G359.89-0.08,
G359.90-0.06, G359.95-0.04, G359.983-0.046, and G0.13-0.11, reported
and studied earlier in the literature. Most of their hydrogen column
density $N_H$ are of the order of $\sim 10^{22}-10^{23}~{\rm
cm}^{-2}$, 
consistent
with other $N_H$ estimates around the GC region. Most of their
photon indices $\Gamma$ fall within the range of $\sim 1-2.5$. The
chance of these X-ray filaments being background extragalactic
sources is very small according to the spectral and morphological
properties. For instance, it would be very difficult to explain the
linear filamentary morphology using the hypothesis of extragalactic
origins. Therefore, these X-ray filaments are most likely unique
objects around the GC region.

As already discussed in Section 1, there were suggestions that these
X-ray filaments may be ram-pressure confined PWNe (Wang et al. 2003,
2006), or synchrotron emissions from MHD shocks associated SNRs or
emissions resulting from inverse Compton scattering (Yusef-Zadeh et
al. 2005; Figer et al. 1999). The
non-thermal X-ray emission mechanisms may be either synchrotron
emission or inverse-Compton scattering. Magnetohydrodynamic (MHD)
relativistic pulsar winds (Michel 1969; Goldreich \& Julian 1970;
Kennel \& Coroniti 1984a, b; Lou 1996, 1998) and MHD shock
interactions of magnetized outflows (e.g., Yu \& Lou 2005; Yu et al.
2006; Lou \& Wang 2006, 2007) with the interstellar medium (ISM) in
SNRs and PWNe could provide high-energy electrons needed in these
two radiation mechanisms (e.g., Sakano et al. 1993; Lu et al. 2003;
Wang et al. 2006). If these X-ray filaments are SNRs, their
elongations would probably represent MHD shock fronts and therefore,
one would not expect to see a tendency of spectral softening along a
filament. The two bright filaments F3 and F10 both show evidence for
such a softening tendency. The fact that most of these filaments
have point-like sources at the heads also againsts the SNR origin. 
On the other hand, as nonthermal X-ray emission is only
detected in several SNRs within the entire Galaxy, it would be
highly unlikely that there are so many nonthermal SNRs around the GC
region. For this reason, we would argue that most of these X-ray
filaments are not SNRs. Observed properties of these X-ray filaments
may be more consistent with those of PWNe. Typical features of a PWN
are: non-thermal X-ray spectrum, with photon index of $1.1-2.4$ and
a X-ray luminosity $L_x$ range from $4\times 10^{32}$ to $2\times
10^{37}~{\rm erg}~{\rm s}^{-1}$ in $0.2-10$ keV band (e.g.,
Gaensler \& Slane 2006; Kaspi,
Roberts \& Harding, 2006). The $\Gamma$ and $L_x$ of these 10
filaments are consistent with the values of a PWN.
The existence of point-like X-ray sources as
indicated by the image study also tends to favor a PWN scenario.

We may estimate the ages of the putative pulsars with the X-ray
luminosities of these X-ray filaments. Li et al. (2007) studied
statistically the nonthermal X-ray emission from young rotation
powered pulsars and PWNe. They noted that there exists a correlation
between the pulsar age $\tau$ and the $2-10$ keV PWN luminosity
$L_{x,pwn}$, which can be expressed as
$L_{x,pwn}=10^{41.7}\tau^{-2.0\pm 0.3}$. The X-ray luminosities of
these 10 filaments are in the range of 0.2-2.2$\times 10^{33}$ erg
s$^{-1}$. Using this empirical formula, ages of these putative
pulsars are possibly between $\sim 10^3$ to $3\times10^5$ yr.
However, given the dispersion about the above empirical relationship
(Li et al. 2007), the estimate may be uncertain probably by a factor
of 10.

Since the ages of pulsars with bright PWNe are usually younger than
a few tens of thousand years, one may doubt if a pulsar at the age
of several $10^5$ years can produce a detectable X-ray nebula.
However, the PWN of a relatively old pulsar can be enhanced in
surface brightness and thus become detectable if the pulsar wind
materials are confined to one direction. PSR B0355+54 is $\sim
5.6\times10^{5}$ yr old. It converts $\sim 1$\% of its spin-down
luminosity to the cometary-like X-ray nebula (e.g.,
Tepedelenlio\v{g}lu \& \"{O}gelman 2007). The old pulsar PSR
B1929+10 ($\tau\sim3\times10^6$ yr) also converts $2.1\times10^{-4}$
of its spin-down luminosity $\sim 3.9\times10^{33}$ erg s$^{-1}$
into the emission of the cometary nebula (e.g., Becker et al. 2006).
The X-ray filaments identified in the GC region are similar to these
two systems and thus probably powered by pulsars.

Now we discuss whether the number of X-ray filaments, if identified
with PWNe, would be consistent with the estimated star formation
rate in the GC region.
According to Figer et al. (2004), the star formation rate at the GC
is about $10^{-7} M_{\odot}\hbox{ yr}^{-1}\hbox{ pc}^{-3}$ which is
some 250 times higher that the mean star formation rate in the
Galaxy. In the field of view of our Fig. 1, we take a radius of
about $7$ pc and estimate the star formation rate to be $1.4\times
10^{-3}M_{\odot}\hbox{ yr}^{-1}$. If the mean mass of a star is 
10$M_{\odot}$, the frequency of supernova explosions would be $1.4\times 10^{-4}$ per
year, leading to about 40 pulsars in the field of Fig. 1 younger
than $\sim 3\times 10^5$ yr as estimated above. This number is
roughly consistent with the 15 candidate PWNe identified in the field.

\subsection{\label{subsec: F10_radio}G0.029-0.06
(F10) and its radio counterpart}

Filament F10 bears certain unique features to be noted here. First,
it has the longest linear structure with the entire image slightly
bent towards the northeast, more or less like an arc. Second, it is
the farthest away from \hbox{Sgr~A$^*$} and thus has much less
contamination from the strong diffuse X-ray emission of Sgr A.
Third, there is an obvious 20cm radio NTF coincident spatially with
X-ray filament F10.

The spectral indices for different regions along filament F10 show
evidence of spectral steepening from the ``head" to ``tail" (see
Table 1). When $N_H$ is fixed at the best fit
value $\sim 7\times 10^{23}~{\rm cm}^{-2}$, the $\Gamma$ values for
the ``head", ``middle", and ``tail" regions are 1.1(1.0, 1.3),
1.5(1.4, 1.6), and 1.8(1.7, 1.9), respectively. This might suggest
an energetic particle flow direction from the southeast (head) to
the northwest (tail). A pulsar moving through the magnetized
interstellar medium seems to give a plausible explanation of this
scenario. Indeed, the morphology of F10 does imply a point source in
the ``head" region. The corresponding point spread function (PSF) at
G0.029-0.06 is an ellipse with a size of $\sim 2\arcsec\times
4\arcsec$. For an updated X-ray versus spin-down luminosity
correlation of rotation powered pulsars, a modified empirical
relation is given by equation (3) of Possenti et al. (2002), namely,
$\log L_{x,(2-10)}=1.34~\log\dot{E}-15.34$ where $L_{x,(2-10)}$ is
the X-ray luminosity in $2-10$keV energy band; using this empirical
relation, we would have a $\dot E\sim 10^{36}~{\rm ergs}~{\rm
s}^{-1}$. Since PSRs J1747-2958 and B1929+10 convert about
 2.5\%  and 2.1$\times10^{-4}$ of their spin-down powers to their
 cometary X-ray nebulae (e.g., Gaensler et al. 2004; Becker et al. 2006),
 the ratio $L_x$/$\dot{E}$ of F10 ($\sim 10^{-3}$) indicates that the
 above estimate for $\dot{E}$ is reasonable.

The arc-like X-ray morphology of F10 and its coincidence with a
radio NTF might be a good indicator of its interaction with
the interstellar magnetic field environment of the GC region (Lang
et al. 1999; Wang et al. 2002b).
Similar to the discussion about G0.13-0.11 by Wang et al. (2002b),
we may estimate the magnetic field strength $B$ in the current
context. First, the lifetime $\tau$ of synchrotron X-ray emitting
particles is given by $\tau\sim (1.3 {\rm~yrs})\epsilon^{-0.5}
B_{\rm mG}^{-1.5}$, where $\epsilon$ is the X-ray photon energy in
unit of keV (a value of 4 keV is adopted here) and $B_{\rm mG}$ is
the magnetic field strength in the filament volume in units of mG. 
The simulations of Bucciantini et al.
(2005)
show that the average flow speed in the tail is about 0.8-0.9 $c$.
For a sustained X-ray linear structure, we estimate by requiring
$\tau \geq L_{obs}/(0.85 c)$. Adopting a characteristic angular
length $L_{obs}$ of $47\arcsec$ ($\sim 2 {\rm pc}$), we thus infer a
magnetic field strength $B\sim 0.3$mG, similar to those in the
bright radio NTFs (e.g, Yusef-Zadeh \& Morris 1987; Lang et al.
1999).

We try to outline a few plausible scenarios in the present context and
discuss relevant aspects qualitatively. Magnetized neutron stars move with
peculiar speeds in the range of a few tens of kilometers per second (a
mean space velocities of $\sim 300-400\hbox{ km s}^{-1}$ for young pulsars;
Hobbs et al. 2005; Faucher-Gigu\`ere \& Kaspi 2006) to well over one thousand
kilometers per second ($\sim 1600\hbox{ km s}^{-1}$) and the surrounding ISM
is generally magnetized. Generally speaking, a typical peculiar velocity of
a neutron star is supersonic and super-Alfv\'enic in a magnetized ISM.
Neutron stars or pulsars have different ranges of surface magnetic field
strengths: $10^9-10^{10}$G for millisecond pulsars in binaries,
$10^{11}-10^{12}$G for a wide range of pulsars, and $10^{14}-10^{15}$G
inferred for several magnetars. Several situations may happen. (1) If
a pulsar does not involve an active pulsar wind, its peculiar motion
through the surrounding magnetized ISM would sustain an MHD bow shock
by its magnetosphere as well as a magnetotail. The faster the pulsar
moves, the more linear the system would appear. This is basically like
a bullet moving through an air supersonically and generating a Mach
cone or wake. Relativistic electrons can be produced at the MHD bow
shock and synchrotron emissions can be generated and sustained at
the same time. (2) In a binary system, the fast wind (say, with a
speed higher than 1000 km s$^{-1}$) from a companion star can blow
towards a spinning magnetized pulsar in orbital motion. Here, the
situations of a companion fast wind blowing across a magnetized
pulsar and a pulsar moving through the ISM with a high speed are
more or less equivalent. Again, an
MHD bow shock and a magnetotail can form in association with the pulsar
system. The stronger the companion wind and the faster the pulsar moves,
the more linear the pulsar system would appear. Relativistic electron
and/or positrons can be generated and sustained to power synchrotron
emissions in the bow shock draped around the pulsar magnetosphere. For
such a system, one might be able to detect the presence of the
companion by various independent means. (3) For a pulsar emitting an
active pulsar wind and with misaligned magnetic and spin axes, spiral
forward and reverse shock pairs can be generated in the relativistic
pulsar wind as a result of inhomogeneous wind and eventually the pulsar
wind is stopped by the ISM through a MHD termination shock (e.g., Lou
1993, 1996, 1998). (4) Case (3) can also happen for a pulsar moving with
a high peculiar velocity through a magnetized ISM. (5) Case (3) can also
happen for a pulsar in binary orbital motion with the companion blows a
powerful wind with a speed higher than 1000 km s$^{-1}$. In both cases
of (2) and (5), the center of mass of the binary system may also move
with a high speed through the ISM. One can further speculate several
possible combinations along this line of reasoning (e.g., Chevalier
2000; Toropina et al. 2001; Romanova, Chulsky \& Lovelace 2005).

There are two clumps (referred to as the east and west clumps
hereafter) of diffuse X-ray emission surrounding filament F10. Although
the west clump is also elongated, it is not called a filament
because it contains many substructures. To
see if these clumps are physically related to F10, we extract
their energy spectra separately (see the middle and bottom panels of
Fig. 6). The fitted parameters are listed in Table 3. 
The much higher absorbing column densities
of the two clumps indicate strongly that they are located farther
away from us than F10 is, while these two clumps themselves are
almost at the same distance (see Table 3).
Moreover, their X-ray emissions are very likely powered by the
same mechanism, as hinted by the characters of their spectra,
which can be fitted well with an absorbed power-law model plus a
6.4 keV emission line. The total emission comes
mostly from the photon energy $4-7$ keV band, with a strong 6.4
keV neutral Fe K line. In contrast, the
Fe line feature is not present in the spectrum of F10. A possible
explanation for this non-thermal, apparently broadened iron line
emission at 6.4 keV is the collisions of low-energy cosmic-ray
electrons with irons in molecular clouds (e.g., Valinia et al.
2000)
or by the radiative illumination from the GC massive black hole
that was suggested to be very bright in the past (e.g., Koyama et
al. 1989).
In conclusion, F10 and the surrounding clumps do not seem to
interact directly.

\subsection{\label{subsec:B-fields}X-ray NTFs as tracers of
the small-scale magnetic fields and gas dynamics}

The oritentation of the X-ray filaments provides an opportunity
to probe the physics conditions of the GC region.
As discussed in section 4.1, the cometary shapes of the X-ray 
filaments imply that the pulsar wind particles are confined to
one direction by the ambient materials.  Mechanism shaping the 
filamentary structure could be ordered magnetic fields 
(e.g., Yusef-Zadeh \& Morris 1987; Lang,
Morris, \& Echevarria 1999) and/or high relative velocity
between the pulsars and the surrounding gas (e.g., Wang et al. 1993; 
Shore \& LaRosa 1999). The magnetic field could be the
product of the gas motion, and the magnetic field could also
control the motion of the gas (e.g., Heyvaerts et al. 1988; Chevalier 1992).
Radio NTFs suggests that the magnetic field is poloidal at large-scale 
 (e.g., Yusef-Zadeh \& Morris 1987; Lang, Morris, \& Echevarria 1999) with 
some more complex smaller structures around the GC region (e.g.,
Nord et al. 2004). 
By looking at the X-ray images shown by Figures 1, 2 and 3, 
there seems to be a tendency that the PWN tails point
away from the GC, indicating that the pulsar wind particles 
are blown outward.  This
might imply the presence of a Galactic wind of hot plasma blowing
away from the center, given the high star formation rate (and so plenty 
of hot gas) in this region. 
Pulsars may have typical peculiar velocities of $\sim
400\hbox{ km s}^{-1}$ (e.g., Hobbs et al. 2005; Faucher-Gigu\`ere \&
Kaspi 2006) and we would expect them to move in random directions.
The tendency for the structures of ten PWNs to orient away from the
center seems to suggest that the Galactic wind has a speed
comparable to or greater than $\sim 400\hbox{ km s}^{-1}$. 

In the above scenario, the particle flow direction of 
the X-ray filament F3 should 
be from the southwest (closer to Sgr A*) to the northeast. This suggests 
that the pulsar, the origin site of the particles, is actually in the 
``tail'' region defined in Fig 2. Then the evident spectral steepening 
from the southwest to the northeast (see section 3) can be naturally 
expained. Therefore, the spectral evolution along F3 also supports the 
existence of a radial high velocity wind in the GC region.

While the X-ray filaments may not completely overlap with their radio
NTFs, their overall orientations are similar. This is supported
by the four X-ray NTFs (including filament F10 in our analysis)
that have radio counterparts: G359.54+0.18 overlays exactly on
the northern part of the two radio filaments (e.g., Yusef et al.
2005; Lu et al. 2003);
G359.89-0.08 and its radio counterpart SgrA-E overlaps partly and
extends in the same direction, with a centroid offset of $\sim 10
\arcsec$ (e.g., Yusef et al. 2005; Lu et al. 2003);
G359.90-0.06 (SgrA-F)
(e.g., Yusef et al. 2005) and G0.029-0.06 (F10) (see our subsection
3.10) also show similar spatial property. Generally speaking, the
X-ray NTFs tend to be shorter than the radio NTFs. This centroid
offset and smaller extent of X-ray filaments could be both explained
by the much shorter synchrotron cooling lifetime in X-ray than in
radio (e.g., Ginzburg \& Syrovatskii 1965).

\section{Summary and Discussion}

To summarize, the most important properties of the GC X-ray filaments
in this {\it Chandra} data analysis are their non-thermal spectra, and
of course, the apparent thread-like linear morphology with point-like
sources residing at the ``head". The scenario of pulsars ploughing
through the interstellar medium and high energy particles from the PWN
interacting with the surrounding magnetic field seems to be a plausible
interpretation. However, this relationship with pulsars can only be
confirmed by the direct detection of periodic flux variations from the
putative point-sources. Future faster timing, higher sensitivity and
spatial resolution of X-ray instruments, together with radio and
near-IR follow-up observations, should help to reveal the true identity
of these X-ray filaments. Meanwhile, these filaments could also provide
some clues to the searching for GC pulsars, as well as the study of
small-scale magnetic field  structures and gas dynamics around the GC.

By our suggestion, one might get the impression that there are many
pulsar bow shocks within $\sim 15$pc of the Galactic center. Actually
in this scenario, there should be even more due to the unavoidable
projection effect; some X-ray bright sources may be heading directly
towards or away from us with their `tails' behind or in front superposed
with their `heads'. We now only catch those suspects readily identified
by their linear or arc-like morphologies. By all astronomical standards,
our Galactic center is not very active at the present epoch even though
there is a black hole with a mass of $\sim 4\times 10^6M_{\odot}$ lurking
at the center near Sgr A$^{*}$ (e.g., Shen et al. 2005). Nevertheless,
the neighborhood of our Galactic center might have been active in the
past with starbursts and magnetized Galactic winds and so forth. While
interacting with winds or flows of the ISM, the compact remnants of these
massive stars (produced during the starburst phase) with a distribution
of peculiar speeds might then have a higher concentration around the
Galactic center. A typical peculiar speed of a few hundred kilometers
per second would be supersonic and super-Alfv\'enic in a magnetized ISM.
With the projection effect taken into account, the physical linear
extension of such a PWN should be jointly determined by the local ISM
flow direction and the direction of the remnant peculiar velocity.

\section*{Acknowledgments}

The authors thank the referee and Editor E. D. Feigelson for
constructive comments leading to a significant improvement of the
manuscript. We thank Q. D. Wang for very helpful discussion, thank C.C. Lang
for sending us the 20 cm radio image before publication. This
research has been supported in part by the Special Funds for Major
State Basic Science Research Projects of China, by the NSFC grant
10573017 at the Institute of High Energy Physics, by NSFC grants
10373009 and 10533020 at the Tsinghua University, by the Tsinghua
Center for Astrophysics, and by the SRFDP 20050003088 and the
Yangtze Endowment from the Ministry of Education at Tsinghua
University. This publication makes use of data products from
the Two Micron All Sky Survey, which is a joint project of the
University of Massachusetts and the Infrared Processing and
Analysis Center, funded by the National Aeronautics and Space
Administration and the National Science Foundation.

{}
\clearpage
\begin{figure}[]
\figurenum{1}
\centerline{{\includegraphics[width=1.0\textwidth]{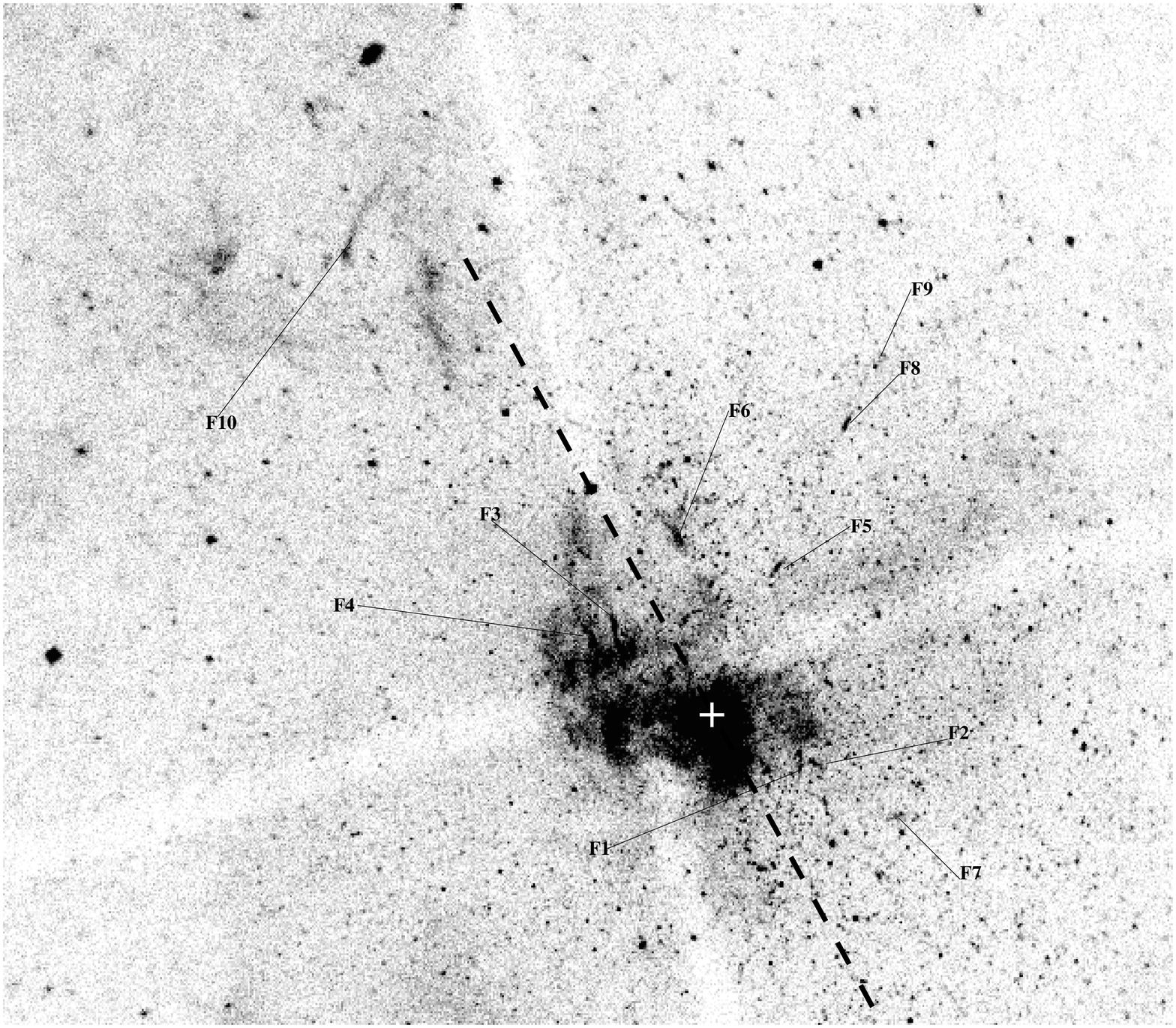}}}
\figcaption[]{A panoramic view of the inner 15 pc around the GC, as
shown by {\sl Chandra} ACIS-I counts in the X-ray photon energy
range of $0.5-8$ keV. The 10 new X-ray filaments analyzed in this
paper are labelled in this image as F1, F2, through F10. Exposure
has not been corrected for this image. The thick dashed line
indicates the direction of the Galactic plane, and the white ``+''
denotes the position of Sgr A*. \label{fig:fig1}}
\end{figure}

\begin{figure}[]
\figurenum{2}
\centerline{{\includegraphics[width=0.9\textwidth]{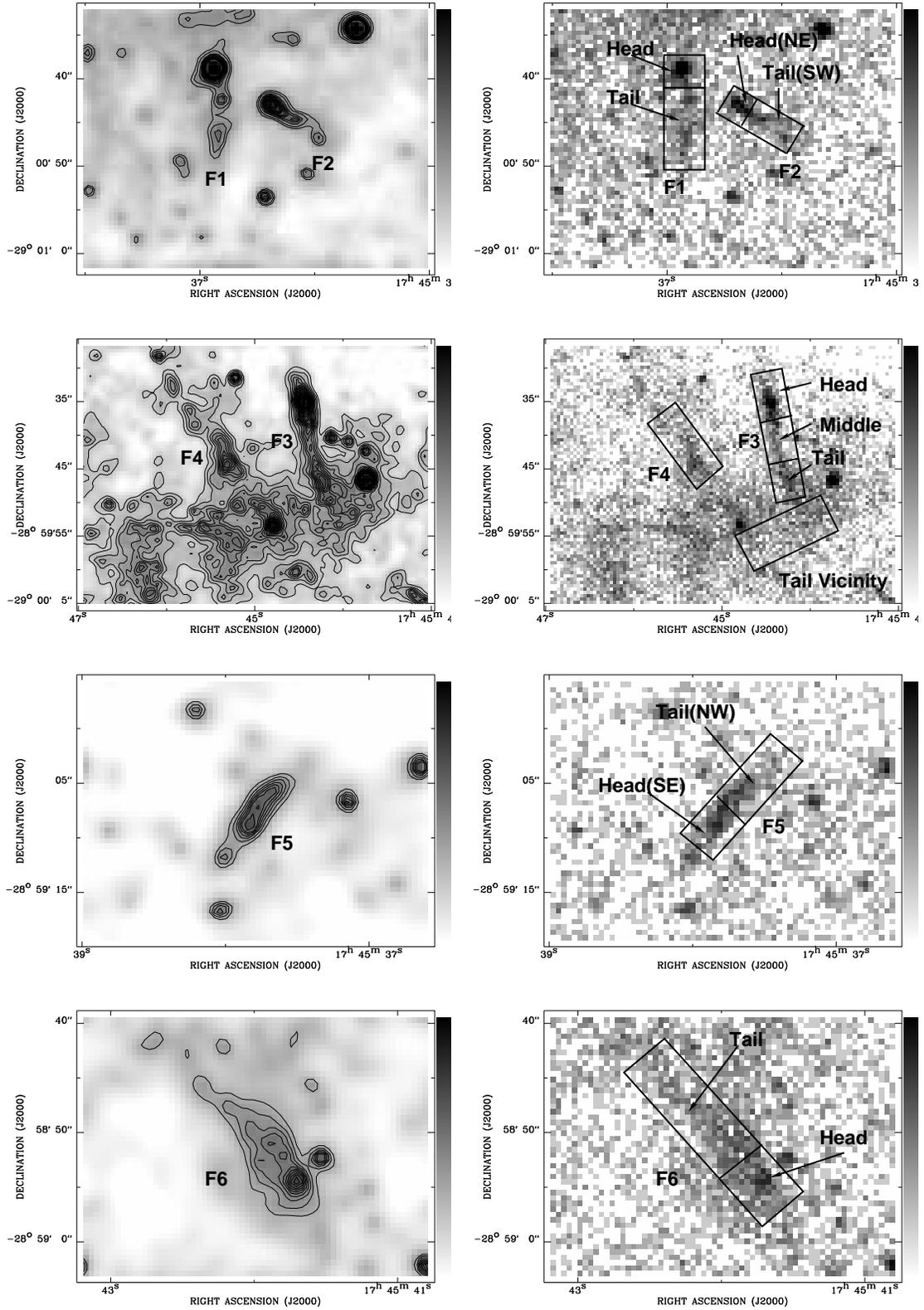}}}
\figcaption[]{
Caption of this Figure 2 is on a separate page -- page 21.
\label{fig:fig2}}
\end{figure}

\vfill\eject
\clearpage
Figure 2 caption here.

{\it Left:} The 0.5-8.0 keV X-ray images of F1 to F6 with contours overlaid. These
images are smoothed adaptively with a Gaussian to achieve a
signal-to-noise ratio of 10. For the same data, these contours
have the same stepsize of 4 counts arcsec$^{-2}$ but different
starting values, i.e., 20 counts arcsec$^{-2}$ for F1 to F4 and
16 counts arcsec$^{-2}$ for F5 and F6.
{\it Right:} The count maps of the same regions as in the left
column. The color bars are in logarithmic scale. For F1, F2, F5
and F6, the lower and upper limits are 4 counts arcsec$^{-2}$ and
120 counts arcsec$^{-2}$, while for F3 and F4 the two limits are
12 counts arcsec$^{-2}$ and 120 counts arcsec$^{-2}$, respectively.
rays are displayed needs to be given in the caption.

\begin{figure}[]
\figurenum{3}
\centerline{{\includegraphics[width=0.9\textwidth]{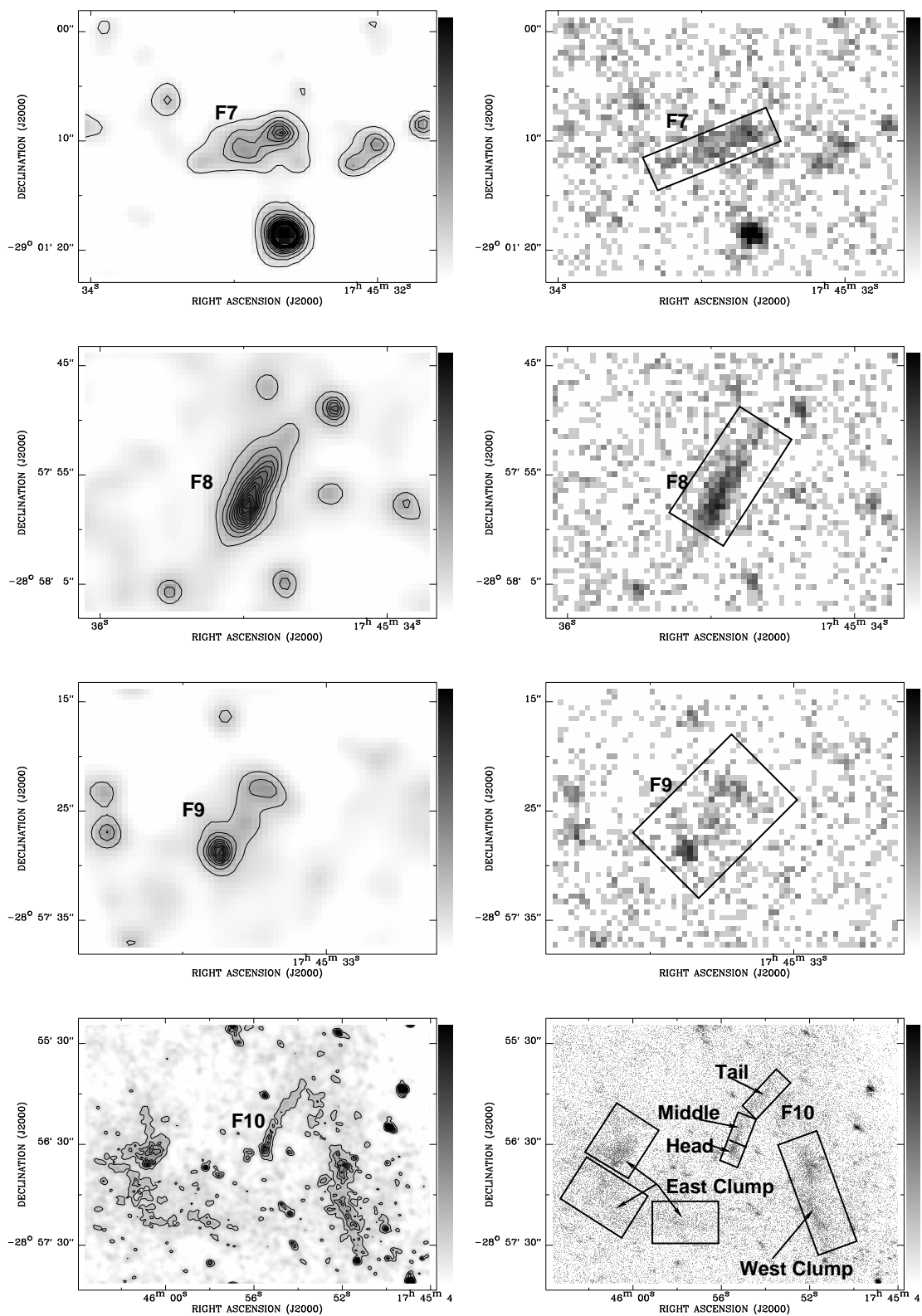}}}
\figcaption[]{
Image arrangement similar to Figure 2 but for four X-ray
filaments F7 to F10. In all plots, the contours start from 8
counts arcsec$^{-2}$ with a stepsize of 4 counts arcsec$^{-2}$.
The grey bars on the right of each image have the same meaning
of those for filaments F1 and F2 as shown in Fig. \ref{fig:fig2}.
\label{fig:fig3}}
\end{figure}

\begin{figure}[]
\figurenum{4}
\centerline{{\includegraphics[width=1.0\textwidth]{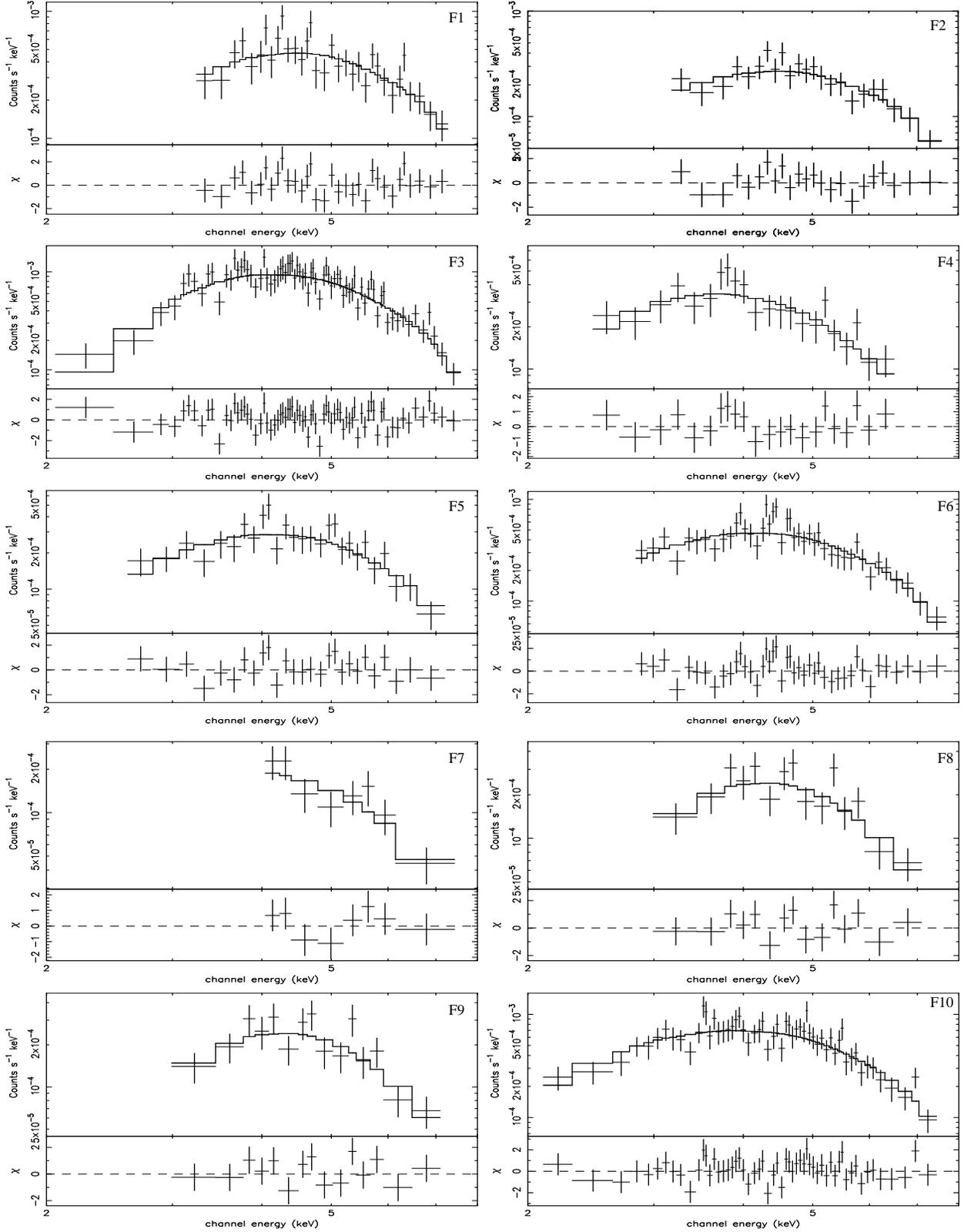}}}
\figcaption[]{The spectra of the ten X-ray filaments around
the GC in the photon energy $2-8$ keV band. All spectra are
fitted with the absorbed power-law model ($phabs(po)$).
\label{fig:fig4}}
\end{figure}

\begin{figure}[]
\figurenum{5}
\centerline{{\includegraphics[width=1.0\textwidth]{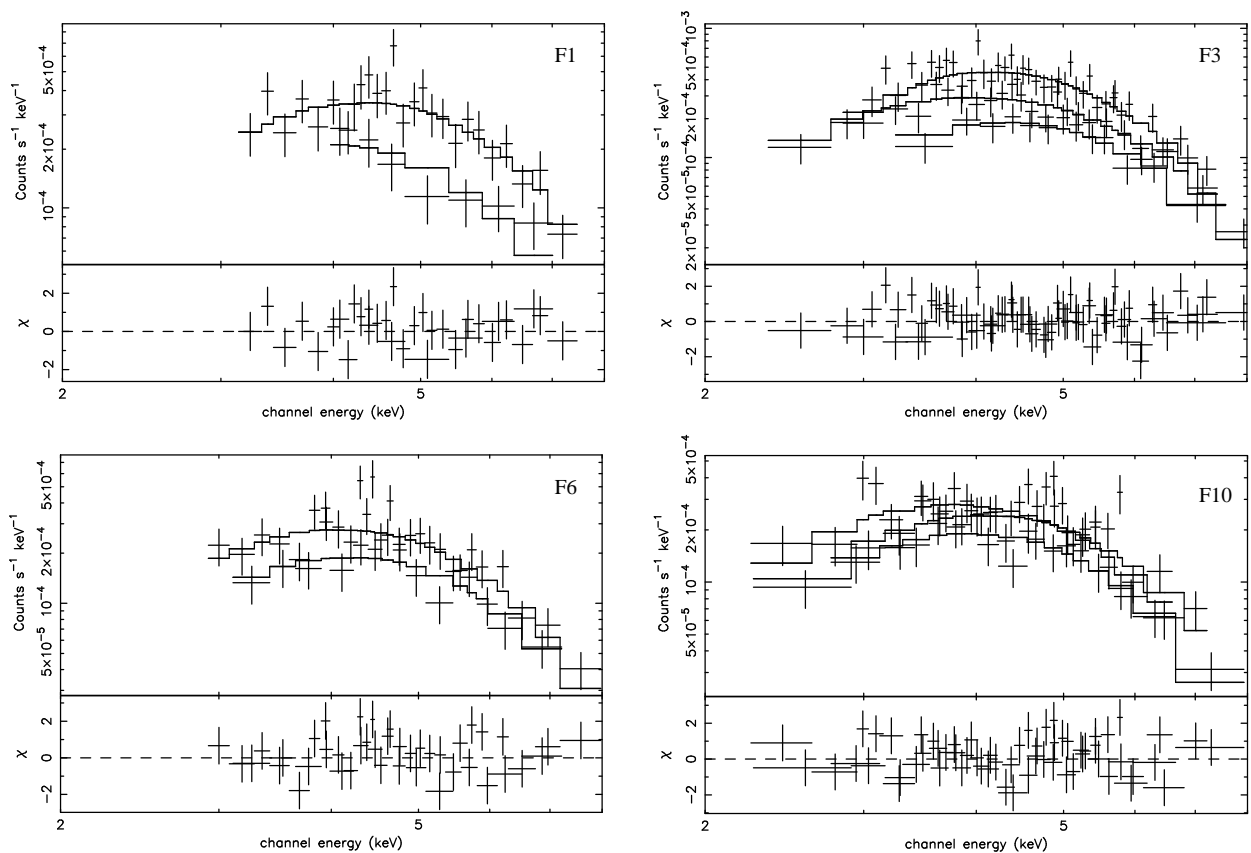}}}
\figcaption[]{The X-ray spectra determined from the sub-regions of
X-ray filaments F1, F3, F6, and F10 in the photon energy $2-8$ keV
band. All spectra here are fitted with the absorbed power-law model
($phabs(po)$). \label{fig:fig5}}
\end{figure}

\begin{figure}[]
\figurenum{6}
\centerline{{\includegraphics[width=0.5\textwidth]{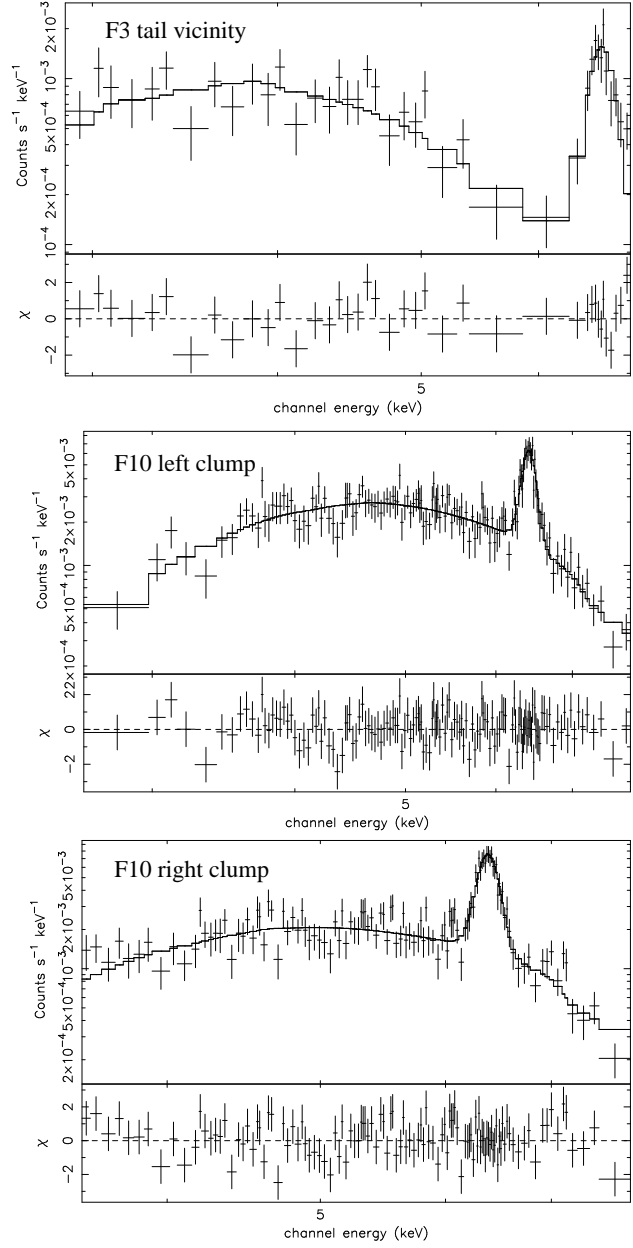}}}
\figcaption[]{
The X-ray spectra of a vicinity region close to the tail of filament
F3 ({\it top panel}), the left (East) clump ({\it middle panel}) and
the right (West) clump ({\it bottom panel}) surrounding filament
F10.\label{fig:fig6}}
\end{figure}

\begin{figure}[]
\figurenum{7}
\centerline{{\includegraphics[width=0.6\textwidth]{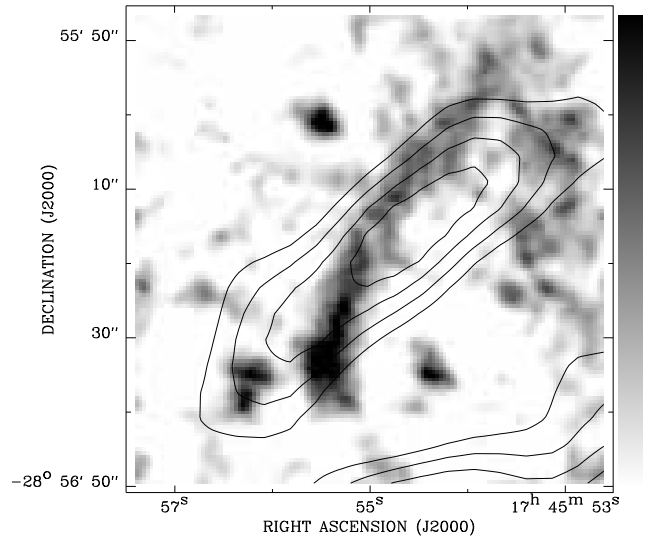}}}
\figcaption[]{
The 20 cm radio contours superimposed onto the
X-ray image of F10. The radio data was obtained by the Very Large Array
(VLA) on July 23, 2001, with a beam size of $15\arcsec\times15\arcsec$.
The contours are at 0.14, 0.15, 0.16 and 0.17 Jansky/beam (Lang et al. 2007).
The X-ray
image is smoothed adaptively so as to achieve a signal-to-noise ratio
of 8, and the greyscale changes from 5 to 20 counts arcsec$^{-2}$
logarithmically.
\label{fig:fig7}}
\end{figure}


\clearpage
\begin{deluxetable}{lccccc}
\tabletypesize{\tiny}
\tablecaption{Spectral fitting results of the
X-ray filaments\label{tab:spect_all}} \tablehead{
Region & $N_H$  & $\Gamma$ &$F_x$&$L_x$& $\chi^2/d.o.f.$\\
&$10^{23}{\rm cm}^{-2}$& &$10^{-14}{\rm erg}~{\rm s}^{-1}{\rm
cm}^{-2}$ &$10^{33}{\rm erg}~{\rm s}^{-1}{\rm d_8}^{2}$ }

\startdata
\multicolumn{6}{c}{X-ray Filament 1 (F1, G359.936-0.038)}\\
Whole   & 1.0(0.2, 1.5)& 0.9(-0.3, 1.6) & 8.0 (7.3, 10.2) &0.92 (0.84, 1.15) &29.8/30  \\
Joint-fit & set $N_H$ the same \\
Head(N) & 0.8(0.3, 1.7)& 0.8(-0.2, 2.0) & && 26.3/28\\
Tail(S) & 0.8(0.3, 1.7)& 1.6(0.1, 3.3)  & && 26.3/28\\

\multicolumn{6}{c}{X-ray Filament 2 (F2, G359.934-0.037)}\\
Whole    & 0.9(0.2, 2.1)& 0.8(0.3, 2.0)  & 4.7 (4.2, 5.3)&0.53 (0.44, 0.97)& 13.2/20 \\
Joint-fit& set $N_H$ the same \\
Head(NE) & 0.9(0, 2.7)  & 0.8(-0.8, 2.9) & & & 17.1/14\\
Tail(SW) & 0.9(0, 2.7)  & 1.2(-0.6, 3.4 )& & &17.1/14 \\

\multicolumn{6}{c}{X-ray Filament 3 (F3, G359.965-0.053)}\\
Whole   &1.4(1.1, 1.7) & 2.3(1.8, 2.8) & 11.9 (11.4, 12.5) &2.2 (1.9, 2.78) & 72.2/71 \\
Head    &1.9(1.3, 2.6) & 3.1(2.2, 4.3) & && 31/35  \\
Middle  &1.8(0.6, 4.2) & 2.2(0.6, 5.3) & && 3.6/10  \\
Tail    &0.7(0.3, 1.1) & 1.4(0.9, 1.8) & && 23.5/24\\
Joint-fit & set $N_H$ the same \\
Head    &1.2(0.9, 1.5) & 2.0(1.5, 2.7)&&&\\
Middle  &1.2(0.9, 1.5) & 1.5(0.8, 2.2)&&&\\
Tail    &1.2(0.9, 1.5) & 2.4(1.7, 2.8)&&&\\

\multicolumn{6}{c}{X-ray Filament 4 (F4, G359.964-0.056)}\\
Whole   & 0.6(0.1, 1.2)& 1.9(1.2, 3.0) & 4.0 (3.4, 5.2) &0.47 (0.41, 0.75) & 13.8/19 \\
\multicolumn{6}{c}{Filament 5 (F5, G359.959-0.028)}\\
Whole   & 0.8(0.4, 1.5)& 1.4(0.9, 2.5) & 4.3 (3.6, 4.8) &0.52 (0.43, 0.79) & 19.5/24\\
Head(SE)& 0.5(0., 1.4) & 0.9(-0.4, 2.6)&&&  9.9/10\\
Tail(NW)& 0.8(0., 5.9) & 0.8(-1.2, 6.8)& &&  2.7/2 \\
Joint fit:& set $N_H$ the same \\
Head(SE)& 0.6(0.2, 1.5)& 1.1(-0.3, 2.1)&&& \\
Tail(NW)& 0.6(0.2, 1.5)& 0.9(0.2, 1.5) &&& \\

\multicolumn{6}{c}{X-ray Filament 6 (F6, G359.969-0.038)}\\
Whole   & 0.9(0.4, 1.4)& 1.4(0.9, 2.3)  & 6.6 (5.7, 7.3) &0.84(0.69, 1.07) & 35/40 \\
Head(SW)& 0.7(0., 2.5) &  1.1(-0.2, 3.8)&  && 18/13 \\
Tail(NE)& 0.7(0.1, 1.5)& 1.3(0.6, 2.3)  &  && 25.4/23 \\
Joint fit:& set $N_H$ the same \\
Head(SW)& 0.8(0.3, 1.5)& 1.2(0.3, 2.3) & &   \\
Tail(NE)& 0.8(0.3, 1.5)& 1.6(0.7, 2.0) & &   \\

\multicolumn{6}{c}{X-ray Filament 7 (F7, G359.920-0.029)}\\
whole   & 0.(0., 3.6)& 0.9(-0.2, 4.4) & 2.8 (1.6, 3.0) &0.21 (0.19, 1.96) &5.1/5.0 \\

\multicolumn{6}{c}{X-ray Filament 8 (F8, G359.970-0.009)}\\
whole & 1.5(0.6, 3.0)& 2.3(0.8, 4.6)  & 3.0 (2.3, 3.5) &0.6 (0.4, 1.7) &12.4/12 \\

\multicolumn{6}{c}{X-ray Filament 9 (F9, G359.974-0.000)}\\
whole & 0.7(0., 1.5)& 0.6(-0.9, 1.9)  & 4.7 (4.1, 5.1) &0.41 (0.40, 0.46) & 5.8/13 \\\\

\multicolumn{6}{c}{X-ray Filament 10 (F10, G0.029-0.06)}\\
Whole & 0.5(0.4, 0.7)& 1.2(0.9, 1.5)& 10.4 (9.9, 11.1) &1.06 (1.02, 1.16) & 48.9/55\\
Head  & 1.0(0.6, 1.6)&1.8(1.3, 2.8) & & &27/21\\
Middle& 0.4(0., 1.1) &1.0(0., 2.0)  &  &&8.7/13\\
Tail  & 0.3(0., 0.9) &1.0(0.5, 1.5) &  && 22/18\\
\multicolumn{6}{c}{Joint fit: set $N_H$ the same and fit all the parameters}\\
Head    & 0.7(0.4, 1.0)&1.1(0.6, 1.8)  & &  & \\
Middle  & 0.7(0.4, 1.0)&1.5(0.8, 2.2)  & &  & \\
Tail    & 0.7(0.4, 1.0)&1.8(1.4, 2.2)  & &  & \\
\multicolumn{6}{c}{Joint fit: set $N_H$ the same and fix it at a value 0.7}\\
Head    & 0.7& 1.1(1.0, 1.3) & &  & \\
Middle  & 0.7& 1.5(1.4, 1.6) & &  & \\
Tail    & 0.7& 1.8(1.7, 1.9) & &  & \\
\enddata

\tablecomments{The spectral model used here is the
absorbed power-law. Relevant parameters are\\
the hydrogen gas column density $N_H$, X-ray photon index
$\Gamma$, the observed flux in the energy $2-10$ keV band
$F_x$ and the unabsorbed X-ray luminosity $L_x$ in the
energy $2-10$ keV band. For definitions of notations here,
$d_8$ denotes the distance in units of 8 kpc, $d.o.f.$ is
the degree of freedom.
Unless otherwise stated in Table 1 as $Joint~fit$, both $N_H$
and $\Gamma$ are free parameters in the fitting procedure.
}
\end{deluxetable}

\begin{deluxetable}{llccccccl}
\tabletypesize{
\tiny} \tablecaption{A parameter summary for all the X-ray
filaments\label{tab:table_sumup}} \tablehead{
No.&Name&Galactic&Size&Offset&$\Gamma$&$N_H$&$L_x$&Counterparts\\
&&Coordinates&$\arcsec$& & &$10^{23}{\rm cm}^{-2}$ &$10^{33}{\rm
erg}{\rm s}^{-1}d_{8}^{2}$&Radio/IR\\} \startdata
1&$^{a}$&G359.95-0.04&2$\times$8&$8\farcs7$
&1.9 (1.8, 2.1)&1.4 (1.3, 1.5)&10. &no\\
2&F1&G359.936-0.038&5$\times$17&$40\farcs7$
&0.9 (-0.3, 1.6)&1.0 (0.2, 1.5)&$0.9$&no\\
3&F2&G359.934-0.037&3$\times$9&48$\farcs$
&0.8 (0.3, 2.0)&0.9 (0.2, 2.1) &$0.5$&no\\
4&F3&G359.965-0.053&4$\times$12&$1\farcm3$
&2.3 (1.8, 2.8) &1.4 (1.1, 1.7)&$2.2$&no\\
5&F4&G359.964-0.056&4$\times$12&$1\farcm3$
&1.9 (1.2, 3.0)&0.6 (0.1, 1.2)&$0.5$&no\\
6&F5&G359.959-0.028&4$\times$10&$1\farcm4$
&1.4 (0.9, 2.5)&0.8 (0.4, 1.5)&$0.5$&no\\
7&F6&G359.969-0.038&5$\times$16&$1\farcm6$
&1.4 (0.9, 2.3) &0.9 (0.4, 1.4)&$0.8$&no\\
8&F7&G359.920-0.029&4$\times$15&$1\farcm3$
&0.9 (-0.2, 4.4) &0.0 (0.0, 3.6)&$0.2$&no\\
9&$^{b}$Cannonball&G359.983-0.046&\nodata
&$2\farcm3$&1.6 (1.2, 2.2)&1.6 (0, 4.8)&3.1&no\\
10&F8&G359.970-0.009&4$\times$10&$2\farcm7$
&2.3 (0.8, 4.6) &1.5 (0.6, 3.0)&$0.6$&no\\
11&F9&G359.974-0.00&4$\times$7&$3\farcm3$
&0.6 (-0.9, 1.9)&0.7 (0.0, 1.5)&0.4&no\\
12&$^{c}$SgrA-E&G359.89-0.08&8$\times$30
&$4\farcm0$&2.0 (1.5, 2.5)&3.7 (2.5, 5.1)&$16$&yes (radio)\\
13&F10&G0.029-0.06&6$\times$47&$5\farcm1$
&1.2 (0.9, 1.5) &0.5 (0.4, 0.7)&1.1&yes (radio)\\
14&$^{d}$SgrA-F&G359.90-0.06&3$\times$6&$5\farcm3$
&(0.8, 8.2)&(1.7, 7.9)&$10.8$&yes (radio)\\
15&$^{e}$&G0.13-0.11&2$\times$40&$12^\prime$
&1.8 (1.4, 2.5)&0.6 (0.4, 0.9) &3.2&yes (radio)\\
\enddata 
\tablecomments{ Offset denotes the angular distance of the brightest
part of an X-ray filament to the position of \sgra; $L_x$ is the
absorption corrected X-ray luminosity in the energy $2-10$ keV band,
except that of SgrA-F, which is in the energy $2-8$ keV band. Some
parameters of the filaments are obtained from the references $a-e$
listed below. The last column of the table shows whether the
counterparts associated with the
filaments in radio bands have so far been found or not.\\
$^{a}$ Wang, Lu, \& Gotthelf (2006).\\
$^{b}$ Park, et al. (2005).\\
$^{c}$ Yusef-Zadeh, et al. (2005).\\
$^{d}$ Yusef-Zadeh, et al. (2005).\\
$^{e}$ Wang, Lu, \& Lang (2002).\\ }
\end{deluxetable}

\begin{deluxetable}{lccccccc}
\tabletypesize{\tiny}
\tablecaption{Spectral fitting results of the
east and west clumps near X-ray filament F10\label{tab:f10_clumps}}
\tablehead{
Region&$N_{H}$&$\Gamma$&F$_x$&L$_x$&Line E&E.W.&$\chi^2/d.o.f$ \\
&$10^{23}{\rm cm}^{-2}$& &10$^{-13}$ erg cm$^{-2}$ s$^{-1}$&
$10^{33}$erg s$^{-1}$&keV&keV&\\} \startdata E. C. & 1.4 (1.2, 1.6)&
1.0 (0.5, 1.2)&6.1 (5.9, 6.6) & 8.1 (7.5, 8.4)
&6.42 (6.41, 6.43) & 0.75 (0.68, 0.83) & 112/135\\
W. C. & 2.3 (1.6, 2.9)& 1.0 (0.4, 1.6)& 5.4 (5.0, 5.7) & 8.7 (7.8,
10.9)
&6.39 (6.38, 6.40) & 1.07 (0.94, 1.23) & 126/114\\
\enddata
\tablecomments{Spectral Model: $N_{H}$ (power law+gaussian).
Relevant parameters are: $N_H$: hydrogen gas column density;
$\Gamma$: photon index; $F_x$: absorbed flux in $2-10$ keV band;
$L_x$: X-ray luminosity in $2-10$ keV band; $Line E$: line center
energy; $E.W.$: equivalent width of the line; $\chi^2/d.o.f$:
chi-square divided by the number of degree of freedom (d.o.f).
}
\end{deluxetable}


\begin{thebibliography}{}
\bibitem[Baganoff et al.(2003)]{bag03}Baganoff, F. K., Maeda, Y., Morris,
M., Bautz, M. W., Brandt, W. N., Cui, W., Doty, J. P., Feigelson, E. D.,
Garmire, G. P., Pravdo, S. H., Ricker, G. R., \& Townsley, L. K. 2003,
ApJ, 591, 891
\bibitem[Beck & Hoernes (1996)]{beck96} Beck R., Hoernes P.,
1996, Nature, 379, 47
\bibitem[Becker et al. (2006)]{becker06} Becker, W., Kramer, M.,
Jessner, A., Taam, R. E., Jia, J. J., Cheng, K. S., Mignani, R.,
Pellizzoni, A., de Luca, A., Slowikowska, A., \& Caraveo, P. A.
2006, ApJ, 645, 1421
\bibitem[Becklin et al.(1978)]{beck78} Becklin, E. E., Matthews, K.,
Neugebauer, G., \& Willner, S. P. 1978, ApJ, 219, 121
%
\bibitem[Bian & Lou (2005)]{BianLou05}Bian,
F.-Y. \& Lou, Y.-Q. 2005, MNRAS, 363, 1315
\bibitem[Boldyrev \& Yusef-Zadeh (2006)]{Boldyrev06}Boldyrev, S., \& Yusef-Zadeh, F. 2006,
ApJ, 637, L101
\bibitem[Bucciantini (2002)]{Buc2002a}Bucciantini, N.
2002a, A\&A, 387, 1066
\bibitem[Bucciantini (2002)]{Buc2002b}Bucciantini, N.
2002b, A\&A, 393, 629
\bibitem[Bucciantini (2005)]{Buc2005}Bucciantini, N.,
Amato, E., \& Del Zanna, L. 2005, A\&A, 434, 189
\bibitem[Cheng et al. (2004)]{Cheng04}Cheng, K. S.,
Taam, R. E., \& Wang, W. 2004, ApJ, 617, 480
\bibitem[Cheng et al. (2006)]{Cheng06}Cheng, K. S.,
Taam, R. E., \& Wang, W. 2006, ApJ, 641, 427
\bibitem[Chevalier (1992)]{Chevalier1992}Chevalier, R.A.
1992, ApJ, 397, L39
\bibitem[Chevalier (2000)]{Chevalier2000}Chevalier, R. A.
2000, ApJ, 539, L45
\bibitem[Cordes \& Chernoff (1998)]{cordes98}Cordes, J. M.,
\& Chernoff, D. F. 1998, ApJ, 505, 315
\bibitem[Cordes \& Lazio(1997)]{cordes97}Cordes, J. M.,
\& Lazio, T. J. W. 1997, ApJ, 475, 557
\bibitem[Cowin Morris 2006]{Cowin06}Cowin,
B. J., \& Morris, M. 2006, BAAS,
Vol. 38, p.1147

%
\bibitem[Faucher-Giguere Kaspi (2006)]{FK2006}
Faucher-Gigu\`ere, C.-A., \& Kaspi, V. M. 2006, ApJ, 643, 332
%
\bibitem[Ferriere (2001)]{Ferriere01}Ferriere, K. M. 2001,
Reviews of Modern Physics, 73, 1031
\bibitem[Ferriere (2007)]{Ferriere07}Ferriere, K. M. 2007,
EAS Publications Series, 23, 3
\bibitem[Figer et al.(1999)]{figer99}Figer, D. F.,
Kim, S. S., Morris, M., Serabyn, E., Rich, R. M.,
\& Mclean, I. S. 1999, ApJ, 525, 750
\bibitem[Figer et al. (2002)]{figer02}Figer, D.F., Majarro, F.,
Gilmore, D., Morris, M., Kim, S. S., Serabyn, E., McLean, I. S.,
Gilbert, A. M., Graham, J. R., Larkin, J. E., Levenson, N. A., \&
Teplitz, H. I. 2002, ApJ, 581, 258
\bibitem[Figer et al. (2004)]{figer04}Figer, D.F., Rich, R. M.,
Kim, S.S., Morris, M., \& Serabyn, E. 2004, ApJ, 601, 319
\bibitem[Gaensler et al. (2004)]{gaensler04}Gaensler, B. M.,
van der Swaluw, E., Camilo, F., Kaspi, V. M., Baganoff, F. K.,
Yusef-Zadeh, F., \& Manchester, R. N. 2004, \apj, 616, 383
\bibitem[Gaensler Slane (2006)]{gaensler06}Gaensler, B. M.,
Slane, P. O., 2006, ARA\&A, 44, 17
\bibitem[Ginzburg \& Syrovatskii (1965)]{ginzburg65}
Ginzburg, V. L., \& Syrovatskii, S. I. 1965, ARA\&A, 3, 297
\bibitem[Goldreich \& Julian (1970)]{GW1970}
Goldreich, P., \& Julian, W. H. 1970, 160, 971
\bibitem[Gotthelf \& Olbert(2002)]{got02}Gotthelf, E. V.,
\& Olbert, C. M. 2002, in ASP Conf. Ser.271, Neutron Stars
in Supernova Remnants, ed. P. O. Slane \& B. M. Gaensler
(San Francisco: ASP), 171
\bibitem[Gotthelf(2003)]{got03}Gotthelf, E. V. 2003, ApJ, 591, 361
\bibitem[Gotthelf(2004)]{got04}Gotthelf, E. V. 2004, astro-ph/0610376
\bibitem[Han(2007)]{Han07}Han, J. L. 2007, eprint/arXiv:0705.4175
\bibitem[Heyvaerts et al. (1988)]{Hey88}Heyvaerts, J., Norman, C., \& Pudritz, R.E. 1988, ApJ, 330, 718
\bibitem[Ho et al.(1985)]{ho85}Ho, P. T. P., Jackson, J. M.,
Barrett, A. H., \& Armstrong, J. T. 1985, ApJ, 288, 575

\bibitem[Hobbs et al. (2005)]{Hobbs2005}Hobbs, G.,
Lorimer, D. R., Lyne, A. G., \& Kramer, M., 2005, MNRAS,
360, 974
\bibitem[Johnston et al. (2006)]{johnstonetal06}
Johnston, S., Kramer, M., Lorimer, D. R., Lyne A. G.,
McLaughlin, M., Klein, B., Manchester, R. N. 2006,
MNRAS, 373, L6
\bibitem[Kaspi et al. (2006)]{Kaspi2006}Kaspi,
V. M., Roberts, M. S. E., Harding, A. K. 2006,
In: Compact stellar X-ray sources. Edited by Walter Lewin \&
Michiel van der Klis. Cambridge Astrophysics Series, No. 39.
Cambridge, UK, p. 279-339
\bibitem[Kawai et al. (1998)]{Kawaietal98}Kawai, N.,
Tamura, K., \& Shibata, S. 1998, in Neutron Stars and Pulsars
Thirty Years after the Discovery, ed. N. Shibazaki, N. Kawai, S.
Shibata \& T. Kafune (Tokyo: Universal Academy), 449
\bibitem[Kennel \& Coroniti (1984a)]{KC84a}Kennel, C. F.,
Coroniti, F. V. 1984a, ApJ, 283, 694
\bibitem[Kennel \& Coroniti (1984b)]{KC84b}Kennel, C. F.,
Coroniti, F. V. 1984b, ApJ, 283, 710
\bibitem[Koyama et al.(1989)]{koy89} Koyama, K., Awaki, H.,
Kunieda, H., Takano, S., Tawara, Y., Yamauchi, S., Hatsukade, I.,
\& Nagase, F. 1989, Nature, 339, 603
\bibitem[Lang (1999)]{lang992} Lang, K. R. 1999,
Astrophysical Formulae, Volume II, (Springer), 87
\bibitem[Lang, Morris, \& Echevarria (1999)]{lang99}Lang, C. C.,
Morris, M., \& Echevarria, L. 1999, ApJ, 526, 727
\bibitem[Lang et al. (2007)]{la03}Lang, C. C. et al., 2007, in preparation
\bibitem[Larosa et al. (2005)]{la05}LaRosa, T. N.,
Brogan, C. L., Shore, S. N., Lazio, T. J.,
Kassim, N. E., Nord, M. E. 2005, ApJ, 626, 23
\bibitem[Larosa et al.(2000)]{la00}LaRosa, T. N., Kassim,
N. E., Lazio, T. J. W., \&  Hyman, S. D. 2000, AJ, 119, 207
\bibitem[Li \& Lu (2007)]{li06}Li, X. H.,
\& Lu, F. J.  2007, \apj, in preparation
\bibitem[Lou (1987)]{Lou87}Lou, Y.-Q. 1987, ApJ, 322, 862
\bibitem[Lou (1993)]{Lou93}Lou, Y.-Q. 1993, ApJ, 414, 656
\bibitem[Lou (1996)]{Lou96}Lou, Y.-Q. 1996, MNRAS, 279, 129
\bibitem[Lou (1998)]{Lou98}Lou, Y.-Q. 1998, MNRAS, 294, 443
\bibitem[Lou (2001)]{Lou01}Lou, Y.-Q. 2001, ApJ, 563, L147
\bibitem[Lou \& Bai (2006)]{LouBai06}Lou, Y.-Q.,
\& Bai X. N. 2006, MNRAS, 372, 81
\bibitem[Lou \& Wang (2007)]{LouWang07}Lou, Y.-Q., \&
Wang, W. G. 2007, MNRAS, 378. L54  (astro-ph/0704.0223)
\bibitem[Lu et al.(2002)]{lu02}Lu, F. J., Wang, Q. D., Aschenbach,
B., Durouchoux, P., \& Song, L. M. 2002, ApJ, 568, L49
\bibitem[Lu et al.(2003)]{lu03}Lu, F. J.,
Wang, Q. D., \& Lang, C.C. 2003, AJ, 126, 319
\bibitem[Manchester (2003)]{Manchester03} Manchester, R. N. 2003,
www.atnf.csiro.au/research/pulsar/catalogue/
\bibitem[Michel (1969)]{Michel69} Michel, F. C. 1969, ApJ, 158, 727
\bibitem[Morris(1996)]{mo96}Morris, M., \& Serabyn, E. 1996, \araa, 34, 645
\bibitem[Muno et al.(2006)]{muno06} Muno, M. P.,
Bauer, F. E., Bandyopadhyay, \& Wang, Q. D. 2006,
\aj, submitted, astro-ph/0601627
\bibitem[Muno et al.(2007)]{muno07} Muno, M.P., Baganoff, F.K., Brandt, W.N.,
Morris, M.R., \& Starck, J.L. 2007 ApJ, submitted (astro-ph/0707.1907)
\bibitem[Nord et al.(2004)]{nord04}Nord, M. E., Lazio,
T. J. W., Kassim, N. E., Hyman, S. D., LaRosa, T. N.,
Brogan, C. L., \& Duric, N. 2004, AJ, 128, 1646
\bibitem[Park et al.(2005)]{par05}Park, S., et al. 2005, ApJ, 631, 964
\bibitem[Possenti et al. (2002)]{possenti02}Possenti, A.,
Cerutti, R., Colpi, M., \& Mereghetti, S. 2002, A\&A, 387, 993
\bibitem[Romanova et al. (2005)]{Romanova05}Romanova, M. M.,
Chulsky, G. A., Lovelace, R. V. E. 2005, ApJ, 630, 1020
\bibitem[Sakano et al.(2003)]{sak03}Sakano, M., Warwick, R. S.,
Decourchelle, A., \& Predehl, P. 2003, MNRAS, 340, 747
\bibitem[Seward \& Wang(1988)]{se88}Seward,
F. D., \& Wang, Z. R. 1988, ApJ, 332, 199
\bibitem[Shen et al. 2005]{shen05}Shen, Z.-Q., Lo, K. Y., Liang,
M.-C., Ho, Paul T. P., \& Zhao, J.-H. 2005, Nature, 438, 62
\bibitem[Shore \& LaRosa 1999]{sho99}Shore, S.N., \& LaRosa, T.N. 1999, ApJ, 521, 587
\bibitem[Spitkovsky 2006]{Spitkovsky06}Spitkovsky, A. 2006,
ApJ, 648, 51
\bibitem[Tepedelenlio\v{g}lu \& \"{O}gelman(2007)]{te07}Tepedelenlio\v{g}lu, E. \& \"{O}gelman, H. 2007, ApJ, 658, 1183
\bibitem[Toropina et al. 2001]{Toropina01}Toropina, O. D., Romanova,
M. M., Toropin, Yu. M., Lovelace, R. V. E. 2001, ApJ, 561, 964
\bibitem[Valinia et al.(2000)]{val00}Valinia, A., Tatischeff, V.,
Arnaud, K., Ebisawa, K., \& Ramaty, R. 2000, ApJ, 543, 733
\bibitem[Wang et al. (1993)]{wa93}Wang, Q.D., Li, Z.Y., \& Begelman, M.C. 1993, nature, 364, 127
\bibitem[Wang et al.(2002a)]{wa02a}Wang, Q. D.,
Gotthelf, E. V., \& Lang, C. C. 2002a, \nat, 415, 148

\bibitem[Wang, Lu \& Lang(2002b)]{wang02b}Wang, Q. D.,
Lu, F. J., \& Lang, C. C. 2002b, ApJ, 581, 1148
\bibitem[Wang, Lu \& Gotthelfi (2006)]{wanglu06}Wang, Q. D.,
Lu, F. J., \& Gotthelf, E. V. 2006, \mnras, 367, 937
\bibitem[Wang \& Lou (2007)]{WangLou07}Wang, W. G.,
\& Lou, Y.-Q. 2007, ApSS, in press
\bibitem[Wang, Yang, Zhang, Ma, Zhou, Li, Lou, Li (2005)]
{wangetal2005}Wang, X. F., Yang, Y. B., Zhang, T. M.,
Ma, J., Zhou, X., Li, W. D., Lou, Y.-Q., \& Li, Z. W.,
2005, ApJ, 626, L89
\bibitem[Wielebinski (2005)]{Richard}Wielebinski, R.,
2005, Cosmic Magnetic Fields. eds. Richard Wielebinski
and Rainer Beck. Lecture notes in Physics Volume 664, p89
%
%
\bibitem[Willingale et al.(2001)]{will01}Willingale, R.,
Aschenbach, B., Griffiths, R. G., Sembay, S.,
Warwick, R. S., Becker, W., Abbey, A. F., \&
Bonnet-Bidaud, J.-M. 2001, A\&A, 365, L212
\bibitem[Wu \& Lou (2006)]{WuLou06}Wu, Y.,
\& Lou, Y.-Q. 2006, MNRAS, 372, 992
\bibitem[Yu \& Lou (2005)]{YuLou05}Yu, C., \& Lou, Y.-Q.
2005, MNRAS, 364, 1168
\bibitem[Yu, Lou, Bian \& Wu (2006)]{YuEtal06}Yu, C.,
Lou, Y.-Q., Bian, F.-Y., \& Wu, Y. 2006, MNRAS, 370, 121
\bibitem[Yusef-Zadeh et al.(1984)]{yu84}Yusef-Zadeh,
F., Morris, M., \& Chance, D. 1984, \nat, 310, 557
\bibitem[Yusef-Zadeh et al.(2005)]{yusef05}Yusef-Zadeh, F.,
Wardle, M., Muno, M., Law, C., \& Pound, M. 2005, Advances in
Space Research, 35, 107
\bibitem[Zweibel \& Heiles (1997)]{zweibelheiles97}
Zweibel, E. G., \& Heiles, C. 1997, Nature, 385, 131
\end{thebibliography}
\end{document}